\documentclass[twocolumn,secnumarabic,amssymb, floatfix, nobibnotes, aps, prl, superscriptaddress]{revtex4-1}
\pdfoutput=1

\usepackage{float}
\usepackage{graphicx}
\usepackage{amsmath} 
\usepackage{epstopdf}
\usepackage{setspace}
\usepackage[shortlabels]{enumitem}
\usepackage{mathrsfs}
\usepackage{color,soul}
\usepackage[dvipsnames]{xcolor}
\usepackage{dcolumn,amssymb}
\usepackage{comment}
\usepackage[normalem]{ulem}
\newcommand{\revision}[1]{\textcolor{BrickRed}{{#1}}}
\usepackage{dcolumn} 
\usepackage{natbib}
\usepackage{bm} 
\usepackage{float} 
\usepackage{bbm} 
\usepackage{color}
\usepackage{amsfonts}
\usepackage{lmodern} 
\usepackage{amsmath,amssymb}

\usepackage[colorlinks=true,citecolor=blue]{hyperref}
\hypersetup{colorlinks=true,citecolor=blue,linkcolor=red,urlcolor=blue}

\renewcommand{\vec}[1]{\boldsymbol{#1}}
\newcommand{\tens}[1]{\boldsymbol{#1}}
\newcommand{\bnabla}{\vec{\nabla}}

\begin{document}
\title{Spontaneous rotation of active droplets in two and three dimensions}
\author{Mehrana R. Nejad}
\email{mehrana.raeisiannejad@physics.ox.ac.uk}
\author{Julia M. Yeomans}
\affiliation{The Rudolf Peierls Centre for Theoretical Physics, Department of Physics, University of Oxford, Parks Road, Oxford OX1 3PU, United Kingdom}
\begin{abstract}
We use numerical simulations and linear stability analysis to study active nematic droplets, in the regime where the passive phase is isotropic. We show that activity leads to the emergence of nematic order and of spontaneous rotation in both two and three dimensions. In 2D the rotation is caused by the formation of a chiral $+1$ defect at the center of the drop. 
With increasing activity the droplet deforms to an ellipse, and then to a rotating annulus. Growing droplets form extended active arms which loop around to produce holes. In 3D the rotation is due to a disclination which loops away from and back to the surface, defining the rotation axis. In the bulk the disclination loop ends at a skyrmion. Active extensile flows  deform the droplet to an oblate ellipsoid, contractile flows elongate it along the rotation axis. We compare our results on rotation in two-dimensional droplets with experiments on microtubule and motor protein suspensions and find a critical radius $\sim 700 \mu m$ above which the spontaneous rotation gives way to active turbulence. Comparing the simulation parameters with experiments on epithelial cell colonies shows that the crossover radius for cell colonies could be as large as $2 mm$, in agreement with experiments.

\end{abstract}

\maketitle


\section{Introduction}

There are many examples where small aggregates of living particles spontaneously rotate.
In vivo, persistent angular motion of \textit{Xenopus} and \textit{Drosophila} eggs \cite{cetera2014epithelial} has been shown to be essential for proper embryonic development \cite{gerhart1989cortical}. Multicellular human mammary cell (MCF10A) spheroids embedded in an alginate and Matrigel-based extracellular matrix rotate \cite{brandstatter2021curvature} and there is evidence that the correct formation of spheres of epithelial cells surrounding a hollow lumen relies on rotation \cite{tanner2012coherent,chin2018epithelial,wang2013rotational}. Rotation has been observed in vitro in experiments on confined myoblast colonies \cite{guillamat2022integer}, active microtubule networks \cite{suzuki2017spatial}, and bacterial systems \cite{wioland2013confinement,chen2021confinement} and also in unconfined layers of cells \cite{ascione2022collective,heinrich2020size,segerer2015emergence} and bacteria \cite{nakane2021large}.

There is increasing interest in interpreting biological phenomena in terms of the theories of active matter \cite{Wensink14308,PhysRevLett.125.218004,natpap,PhysRevLett.129.098102,geo}. Active matter models in 2D that lead to collective rotation include generalised Potts models \cite{segerer2015emergence}, simulations of rod-like swimmers \cite{lushi2014fluid},  agent-based simulations of self-propelling particles with intrinsic angular velocity \cite{PhysRevLett.119.058002,liebchen2022chiral}, or in the presence of randomly distributed obstacles \cite{sppobs}, annular confinements \cite{PPR:PPR419377,Narayana}, and activity gradients \cite{copenhagen2018frustration}. 
Continuum simulations have also been able to predict collective rotation in particular geometries. These include circular confinements with anchoring conditions that impose net topological charges \cite{PhysRevE.97.012702,PhysRevLett.109.168105}, colonies with an imposed motility difference between the center and the rim \cite{heinrich2020size}, and polar systems 
\cite{PhysRevLett.92.078101}.  In the presence of no-slip and anchoring boundary conditions in circular confinement, it has been shown that the imposed $+1$ defect leads to the formation of circulating currents, and when activity is larger than a threshold, it leads to oscillatory behaviour of the defects \cite{PhysRevLett.109.168105}.

In 3D, continuum simulations have identified rotation in confined 3D nematic droplets with strong passive anchoring \cite{PhysRevX.9.031051}, in chiral active nematics \cite{carenza2019rotation}, and in active polar systems moving on curved surfaces \cite{glentis2022emergence,PhysRevX.7.031039,glentis2022emergence}. Agent-based simulations of self-propelling particles moving on a fixed sphere have also reported the formation of collective rotation \cite{chen2019three,PhysRevE.91.022306}. 

Continuum theories of active nematics have been particularly successful in describing the properties of microtubule motor mixtures, bacterial and epithelial cell monolayers \cite{PhysRevLett.127.148001,Wensink14308,saw2017topological,alltogether,doo2018,needleman2017active,xi2019material,prost2015active,Ramaswamy_2017,juelicher2007active,juelicher2007active}. Here we describe how this approach can predict spontaneous rotation in cell aggregates in both two- and three-dimensions. We work in the limit, relevant to systems of isotropic cells, where there is no nematic ordering in the absence of activity \revision{\cite{Santhosh, vafa2021fluctuations}}, and the boundaries at the edge of the droplets have no imposed confinement or anchoring.

In 2D we find a spontaneous transition to a rotating state above an activity threshold. The rotation is driven by active shear flow due to director misalignments which result from the formation of a +1 or two +1/2 topological defects at the centre of the droplet. Increasing activity elongates the droplets and can lead to lumen (hole) formation at the droplet centre.

We also observe rotation in 3D droplets, with a disclination loop forming along the axis of rotation. The resultant active flows extend the droplet along (perpendicular to) the rotation axis for extensile (contractile) active forcing.

We first introduce the equations of motion. We then perform a linear stability analysis to find the onset of collective motion and explore the phase diagram of 2D active isotropic droplets in terms of surface tension, droplet radius, and activity. We also discuss the role of growth in the emergence of the different dynamical steady states. In the next section of the paper, we study the behaviour of active isotropic droplets in 3D in the absence and presence of growth. Finally, we discuss our results and compare them to experiments. 

\section{Equations of Motion} 
To investigate the dynamical behaviour of colonies in two and three dimensions
we solve the continuum equations of motion for an active
nematic droplet. 
The relevant hydrodynamic variables are an orientational order
parameter $\tens{Q}= \mathcal{S} d \left( \vec{n} \vec{n}-\tens{I}/d \right) /(d-1)$, which describes the magnitude $\mathcal{S}$ and direction
of the nematic order $\vec{n}$ in $d$ dimensions, the concentration of the active material $\phi$, and the velocity $\vec{u}$ ~\cite{DeGennesBook}.

The dynamics of the nematic tensor is governed by ~\cite{beris1994thermodynamics}
\begin{align}
\left(\partial_t + \vec{u}\cdot\bnabla \right) \tens{Q} - \tens{S} &= \gamma \tens{H},
\label{eqn:lc}
\end{align}
where $\gamma$ is the rotational diffusivity and $\tens{S}$ is the co-rotational advection term that accounts for the impact of the strain rate $\tens{E}=(\bnabla\vec{u}^{T}+\bnabla\vec{u})/2$ and vorticity $\tens{\Omega}=(\bnabla\vec{u}^{T}-\bnabla\vec{u})/2$ on the director field.
The co-rotational advection has the form
\begin{align}
\tens{S} &= \left(\lambda \tens{E} + \tens{\Omega}\right)\cdot(\tens{{Q}}+\frac{\tens{{I}}}{d}) + (\tens{{Q}}+\frac{\tens{{I}}}{d})\cdot\left(\lambda \tens{E} - \tens{\Omega}\right) \nonumber\\
&  - (d-1) \lambda (\tens{{Q}}+\frac{\tens{{I}}}{d})\left( \tens{Q} : \bnabla \vec{u}\right),
\label{eqn:cor}
\end{align}
where the flow-aligning parameter $\lambda$ controls the coupling between the orientation field and the flow, determining
 whether the nematogens align or tumble in a shear flow.
The relaxation of the orientational order is described by a free energy $\mathcal{F}=\int f_Q dV$ through the molecular field,
\begin{align}
\tens{H} &= -(\frac{\delta f_Q}{\delta\tens{Q}} - \frac{1}{3}\tens{I} \; \text{Tr}\frac{\delta f_Q}{\delta\tens{Q}}). \label{eqn:molpot}
\end{align}

The nematic free energy density is
\begin{align}
f_Q &= \frac{\mathcal{A}}{2} \tens{Q}^2 + \frac{\mathcal{B}}{3} \tens{Q}^3 + \frac{\mathcal{C}}{4} \tens{Q}^4+ \frac{K_Q}{2} |\bnabla \tens{Q}| ^2. \label{fef}
\end{align}
We choose $\mathcal{A}$, $\mathcal{B}$, $\mathcal{C}$, so that the passive system is in the isotropic phase in equilibrium. In active systems in experiment, one can look at the magnitude of the nematic order in the absence of activity to infer if free energy favours the isotopic or nematic phase. For example, it has been observed in the experiments performed in \cite{hardouin2019reconfigurable} that, without activity, microtubules form an isotropic phase, and the presence of activity leads to the formation of nematic order. Another approach is to use machine learning to find the values of the parameters related to the bulk free energy. This has been used for a microtubule motor protein system in Ref. \cite{PhysRevLett.129.258001} where it was again shown that the ground state of the free energy favours an isotropic phase. This regime is also relevant to epithelial cells, such as MDCK strains, which are, on average, isotropic.

The final term in Eq. \ref{fef} encodes the elastic free energy density due to spatial inhomogeneities in the nematic tensor. We assume a single Frank elastic constant $K_Q$ \cite{DeGennesBook}. 

The concentration field $\phi$ defines the position of the 
 nematic droplet, with $\phi=1$ corresponding to the active phase and $\phi=0$ to the passive phase. 
It evolves according to \cite{cahn1958free}
\begin{align}
\partial_t \phi & + \bnabla \cdot (\vec{u} \phi)= \Gamma_{\phi} \nabla^2 \mu +k_d \phi, \label{eqn:phi}
\end{align}
where the mobility $\Gamma_{\phi}$ quantifies
how fast the concentration field responds to gradients in the chemical potential $\mu =  \frac{\partial f_{\phi}}{\partial \phi} -\bnabla \cdot (\frac{\partial f_{\phi}}{\partial \bnabla \phi})$, and the last term is a source term which describes droplet growth. The concentration free energy density is
\begin{align}
f_{\phi} &= 
 \frac{K_{\phi}}{2} |\bnabla \tens{\phi}| ^2  + \frac{A}{2} (\phi-1)^2 \phi^2 
\end{align}
where $K_{\phi}$ and $A$ control the surface tension and interface width.

\begin{figure*}[t] 
    \centering
    \includegraphics[width=0.99\textwidth]{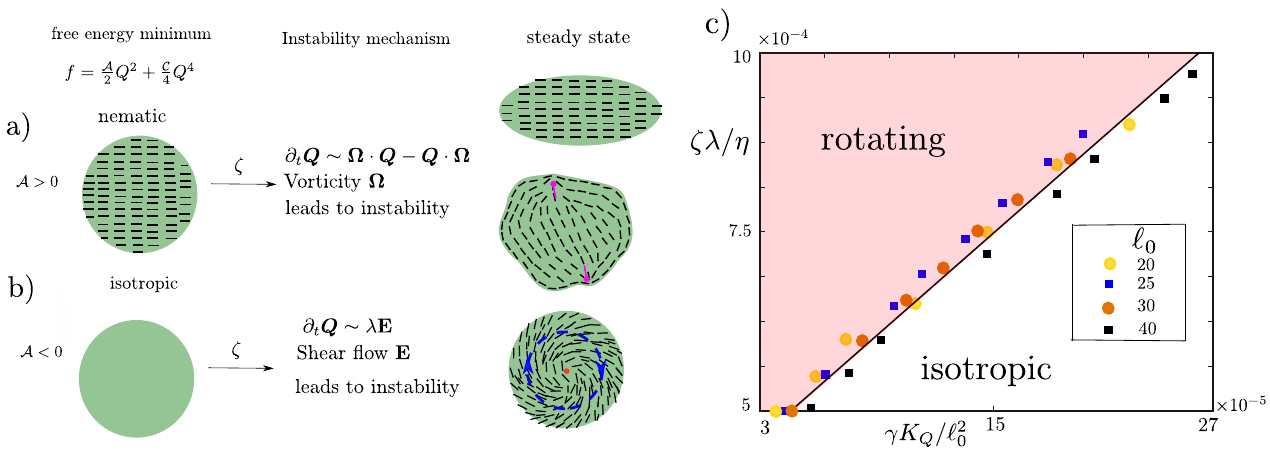}
    \caption{a) In droplets that are in the nematic phase ($S_0\sim 1$), when the elastic constant is large, the droplet elongates (top) \cite{PhysRevLett.113.248303,dell2018growing}. The direction of elongation is along the nematic order in extensile systems and perpendicular to that in contractile systems. For smaller elastic constants, vorticity $\boldsymbol{\Omega}$ destabilises the nematic phase and leads to the formation of defects. The defects lead to interfacial instabilities \cite{geo,dell2018growing}. b) In systems in which the free energy favors an isotropic phase, strain rate $\boldsymbol{E}$ leads to the creation of nematic order and instability of the isotropic phase. When confined to a droplet, activity leads to spontaneous rotation without the need for imposing any topological charge at the interface \cite{ascione2022collective}.  The blue dashed line with arrows shows the direction of rotation of the droplet. c) Emergence of spontaneous rotation in active isotropic droplets in the plane of activity $\zeta$ and scaled elastic constant $K_Q/\ell_0^2$ where $\ell_0$ is the droplet radius. The black line is a linear fit to the data. Here, we have chosen a large surface tension to avoid droplet deformation. }
\label{fig1}
\end{figure*}

\begin{figure*}[t] 
    \centering
    \includegraphics[width=0.99\textwidth]{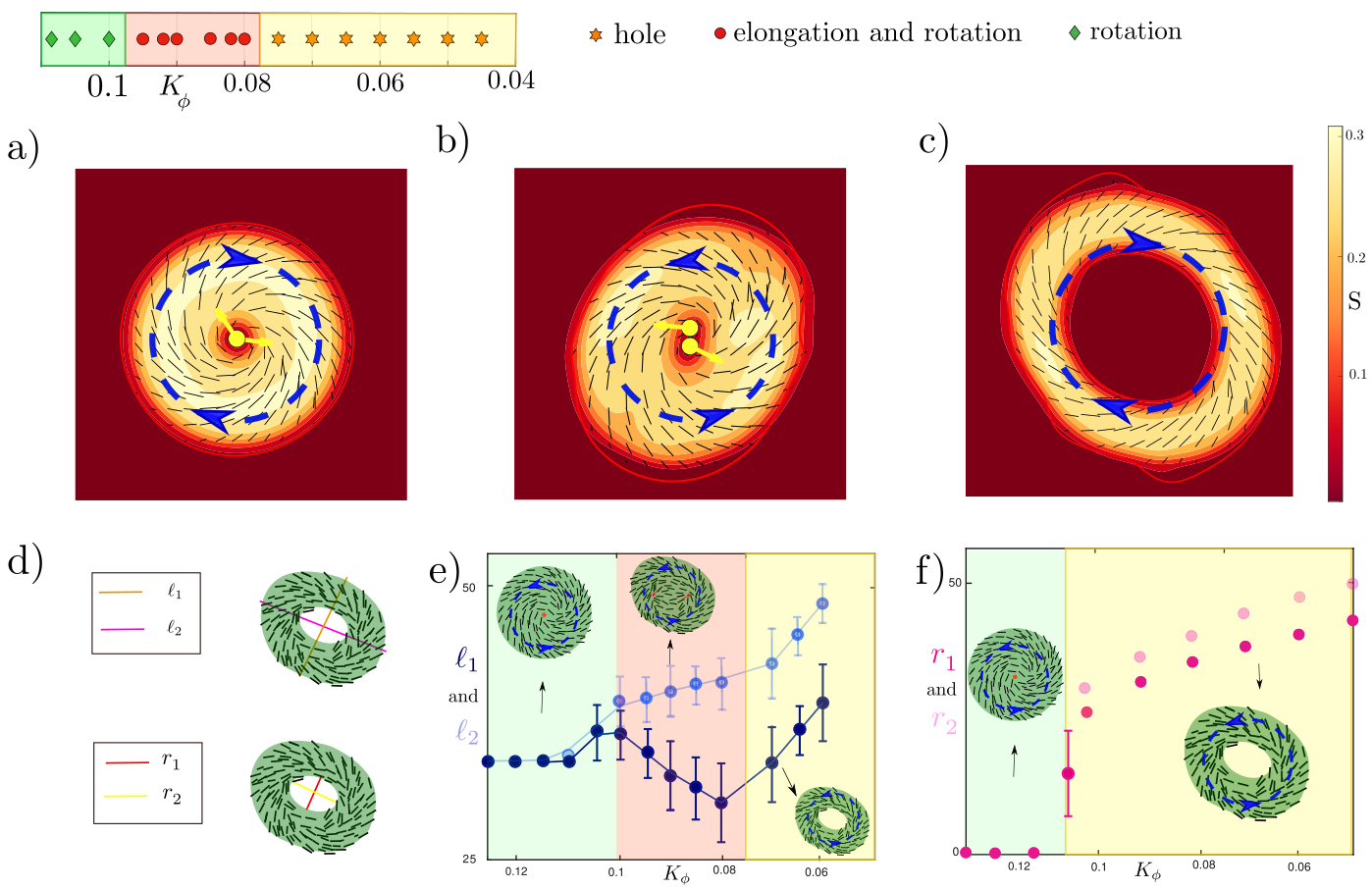}
    \caption{a) Dynamical steady states of a 2D droplet as a function of the surface tension parameter $K_{\phi}$. For large $K_{\phi}$ a circular droplet is stable and undergoes spontaneous rotation (green diamonds). Upon decreasing $K_{\phi}$, the central $+1$ defect splits into two $+1/2$ defects, and the circular droplet deforms into an ellipse  (red circles). Decreasing $K_{\phi}$ further, the droplet forms a hole and the $+1$ defect disappears (yellow stars). The simulations are started from a homogeneous concentration which leads to the formation of a droplet with radius $\ell_0=70$. A two-dimensional phase diagram as a function of elasticity $K_Q$ and surface tension coefficient $K_\phi$ is shown in Fig. S1(a) in the SI. b) $\ell_1$ and $\ell_2$ ($r_1$ and $r_2$) are defined as the smaller and the larger outer (inner) radius of the droplet, respectively.  The remaining panels demonstrate the hysteretic nature of the transitions: in (c) the initial condition (for each data point) is a droplet of radius $70$; in (d) the initial condition is an annulus. The blue dashed lines with arrows show the direction of rotation of the droplet.}
 \label{fig2}
\end{figure*}

\begin{figure}[t] 
    \centering
    \includegraphics[width=0.49\textwidth]{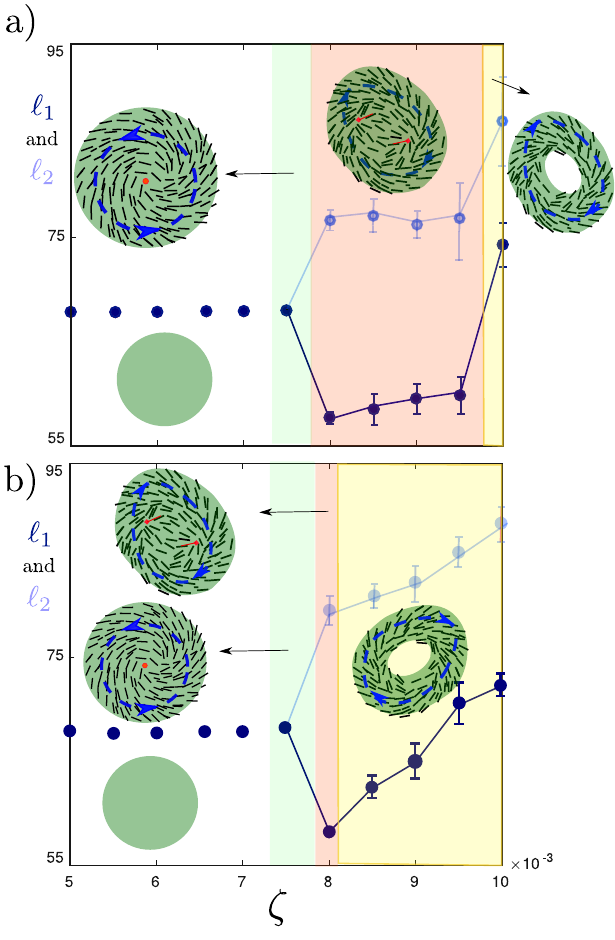}
    \caption{Hysteretic nature of the transitions as activity is varied: in (a) the initial condition is a droplet of radius $70$ (for each data point); in (b) the initial condition is an annulus with the same amount of active material and inner radius $40$.   $\ell_1$ and $\ell_2$ are the smaller and the larger outer  radius of the droplet, respectively. In both (a) and (b)  $K_{\phi}=0.05$. See Supplemental Material at \cite{smnejad} for all the parameters. The blue dashed lines with arrows show the direction of rotation of the droplet.} \label{fig3}
\end{figure}

The dynamics of the velocity field is governed by the incompressible Navier-Stokes equations, which read: 
\begin{align}
\rho \left(\partial_t + \vec{u}\cdot\bnabla\right) \vec{u} &= \bnabla\cdot\tens{\Pi},\label{eqn:cont}\\
\bnabla\cdot\vec{u} &=0. \label{eqn:comp}
\end{align}
Here, $\rho$ is the density and 
$\tens{\Pi}$ is a generalized stress tensor that has both passive and active contributions. The passive part of the stress includes the viscous stress, $\tens{\Pi}^\textmd{visc} = 2 \eta \tens{E}$, elastic stress, 
\begin{align}
\tens{\Pi^{\text{elastic}}} &= -P\tens{I} +2 \lambda \tens{{Q}} (\tens{Q}:\tens{H}) -\lambda \tens{H}\cdot\tens{{Q}}  - \lambda \tens{{Q}} \cdot \tens{H} \nonumber\\
                            &\quad -\tens{\nabla}\tens{Q} : \frac{\delta f}{\delta \tens{\nabla}\tens{Q}} + \tens{Q}\cdot\tens{H} - \tens{H}\cdot\tens{Q},
\label{eqn:elastic}
\end{align}
and capillary stress due to the inhomogeneous concentration field,
\begin{align}
\tens{\Pi}^\text{capillary}= (f_{\phi}+f_{Q}-\mu \phi) \tens{I} -\bnabla \phi (\frac{\partial f_{\phi}}{\partial \bnabla \phi}). \label{eq10a}
\end{align}
In the equations for the passive stress, $P$ is the isotropic pressure, and $\eta$ is the viscosity~\cite{beris1994thermodynamics}. 

The active stress drives changes in the flow field caused by continuous energy injection at the microscopic scale. 
The  activity generates flows for nonzero divergence of the nematic tensor, and the active stress takes the form~\cite{Sriram2002}
\begin{align}
\tens{\Pi^{\text{act}}} &= -\zeta \tens{Q} \phi.
\label{eqn:active}
\end{align}
The parameter $\zeta$ determines the strength of the activity. Extensile (contractile) activity is represented by $\zeta>0$ ($\zeta<0$).
It has been previously shown that phase field models and vertex models, which explicitly resolve individual cells, show a similar behaviour to continuum models, such as formation and annihilation of defects, and velocity and director correlations that decay over a scale much larger than cell size if active nematic driving is added (see \cite{barton2017active,PhysRevLett.122.048004,PhysRevLett.130.038202}), and very recent work is aiming to investigate the mapping between the models in more detail \cite{PhysRevLett.130.058202,underpre}. 

The equations of active nematohydrodynamics are solved using a hybrid lattice Boltzmann and finite difference method~\cite{marenduzzo2007steady,nejad2020memory}, with the discrete space and time steps defining the simulation units.
See Supplemental Material at \cite{smnejad} for the list of parameters and initial conditions. For all the simulations, we use the same set of parameters unless otherwise stated.

{\section{TWO DIMENSIONS}}
\noindent
{\it Linear stability analysis:}

As a first step towards understanding the behavior of the isotropic droplet, we perform a linear stability analysis around an inert isotropic phase, $\textbf{Q}=0+\boldsymbol{\delta} \textbf{Q}$ and $\textbf{u}=0+\boldsymbol{\delta} \textbf{u}$, and study the growth-rate of perturbations in $\boldsymbol{\delta} \textbf{Q}$ and $\boldsymbol{\delta} \textbf{u}$. Since we are working in a low Reynolds number regime, we use an overdamped approximation and ignore density fluctuations. We also consider a free energy cost for the formation of the nematic phase. 
Introducing the Fourier transform for any fluctuating field $f$ as $f(\textbf{r},t)=\int d\omega\: d \textbf{q} \:  \tilde{f}(\textbf{q},t)\: e^{i (\textbf{q} \cdot \textbf{r}+\omega t)}$, Fourier transforming Eqs.~(\ref{eqn:lc},\ref{eqn:cont}-\ref{eqn:comp}) gives
\begin{align}
&i \omega \delta \tilde{Q}_{xx} -(B_1+\frac{\zeta \lambda \sin^2 2\theta}{2 \eta})\delta \tilde{Q}_{xx} - B_2 \delta \tilde{Q}_{xy}=0,\nonumber\\
&i \omega \delta \tilde{Q}_{xy} -(B_1+\frac{\zeta \lambda \cos^2 2\theta}{2 \eta})\delta \tilde{Q}_{xy}-B_2 \delta \tilde{Q}_{xx}=0,
\label{dispersiondy}
\end{align}
where the wave-vector $\textbf{q}=q (\cos{\theta}, \sin{\theta})$, $B_1=\gamma(\mathcal{A} +K_Q q^2)$, and $B_2=(\zeta \lambda \sin 4\theta)/(4 \eta)$.
From Eq.~(\ref{dispersiondy}) the growth rate of the perturbations is
\begin{align}
\Im(\omega_1) &=-\gamma(\mathcal{A}+K_Q q^2),\label{dispersion}\\
\Im(\omega_2) &=\frac{\zeta \lambda}{2 \eta}-\gamma(\mathcal{A}+K_Q q^2). \label{dispersionb}
\end{align}
These equations show that nematic perturbations are suppressed by elasticity $K_Q$ and the free energy term $\mathcal{A}$ that favor an isotropic phase. Conversely, $\Im(\omega_2)$ also includes a term, depending on activity $\zeta$, that destabilizes the isotropic phase. We note that this differs from the instabilities of the isotropic phase in a regime where friction is dominant over viscosity and viscous flows can be ignored \cite{srivastava2016negative}. In particular, in the frictional regime activity appears as a second-order term ($\zeta q^2$) in the growth rate, whereas, in our system, activity appears as a first-order term $\zeta q^0$.

For an active system confined to a droplet, the largest wave vector corresponds to the inverse of the droplet size $q \sim 1/\ell_0$.
Fig.~\ref{fig1}(c) displays simulation results for the onset of nematic ordering and rotation in active droplets showing the expected collapse of the data to a line predicted by Eq.~(\ref{dispersionb}) when plotted in terms of $K_Q/\ell_0^2$ for fixed $\mathcal{A}$. 
 At the linear level, the velocity inside the droplet is given by
\begin{align}
\boldsymbol{\delta} \boldsymbol{\tilde{u}} &= \frac{i \zeta}{\eta q^2} (\frac{\textbf{q} \cdot \boldsymbol{\delta} \boldsymbol{\tilde{Q}} \cdot \textbf{q} }{q}- \textbf{q} \cdot \boldsymbol{\delta} \boldsymbol{\tilde{Q}}).\label{dropvel}
\end{align}
 This shows that once perturbations in the nematic order grow and the nematic order parameter becomes non-homogeneous, the velocity inside the droplet
becomes non-zero.
 Eq.~(\ref{dispersionb}) shows that perturbations in the nematic tensor and the rotational flows grow due to shear, through the tumbling parameter $\lambda$. In Fig.~S3 in the Supplemental Material \cite{smnejad}
we confirmed this by performing simulations where we ignored the vorticity terms $\boldsymbol{\Omega} \cdot \textbf{Q} - \textbf{Q} \cdot \boldsymbol{\Omega}$, and only considered the strain rate $\textbf{E}$. The graphs show that without vorticity the angle between the director and the radius of the droplet is equal to $\pi/4$.  This means that there is a spiral defect at the center of the droplet. The presence of vorticity decreases (increases) the angle in contractile (extensile) systems as contractile (extensile) activity produces splay (bend) when nematic order is formed. 

In the linear stability analysis presented here, we have ignored the effect of the instabilities that can appear in the concentration field. The active instabilities that can form at the interface of droplets are also interesting and have been studied using numerical simulations in ~\cite{xu2023}, and using analytical calculations in \cite{alert2022fingering}. Interface instabilities do not happen at a linear level when there is no anchoring at the interface and $S_0=0$ \cite{alert2022fingering}. In agreement with this, we do not see any interface instabilities before the formation of the order and rotation in our simulations. 
\begin{figure*}[t] 
    \centering
    \includegraphics[width=0.99\textwidth]{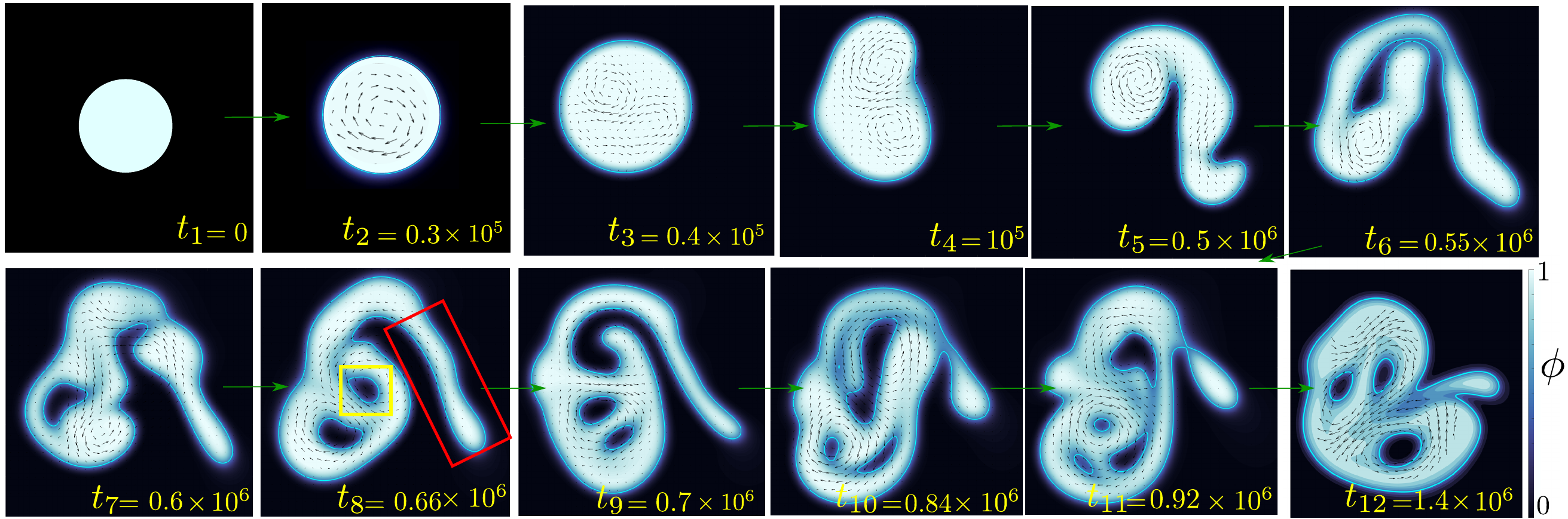}
    \caption{Dynamics of an isotropic growing droplet. The blue background color shows the concentration of the active material and the arrows show the velocity field. The holes (e.g.~yellow outline in $t_8$) result from the formation of isotropic arms (e.g.~red outline in $t_8$) that fold. The green arrows show the direction of time. The snapshots are zoom-in views. Here, we used $K_{\phi}=0.1$. A two-dimensional phase diagram as a function of activity $\zeta$ and surface tension coefficient $K_\phi$ is shown in Fig. S1(b) in the SI.
    \label{growth2d}
    }
\end{figure*}

In our droplet with an isotropic ground state $\textbf{Q}$ remains small, and the active dynamics of the nematic tensor can be approximated by $\partial_t \textbf{Q} \sim \lambda \textbf{E}$ (see Eq. \ref{eqn:cor}). In an isotropic droplet, the instability is caused by the strain rate that leads to formation of a nematic phase and a spiral defect and spontaneous rotation. We would like to emphasise that formation of spontaneous rotation can only be observed in isotropic droplets. Unconfined droplets that are in a nematic phase without activity either elongate, or form active turbulence and interface instabilities \cite{geo,PhysRevLett.113.248303}.

\noindent
{\it Dynamical steady states:}

Beyond the linear regime, surface tension becomes important in controlling droplet behaviour, and the drop shape becomes more complex. The possible steady-state droplet configurations are illustrated in Fig.~\ref{fig2} for varying surface tension $K_{\phi}$ for fixed activity $\zeta=0.01$ and an active area equivalent to $\ell_0 \approx 70$  for the case of a single circular droplet.

As expected, for large values of the surface tension droplets remain circular and active flows form a chiral $+1$ defect at the center of the droplets and lead to spontaneous rotation (Movie 1, and Fig.~\ref{fig2}(a), green diamonds). 
As the surface tension is decreased the droplet deforms to a rotating ellipse and 
the chiral $+1$ defect splits into two $+1/2$ defects (Movie 2, and Fig.~\ref{fig2}(a), red circles).

A further decrease in surface tension shifts the steady state to a droplet with a hole at its center (Fig.~\ref{fig2}(a), yellow stars). 
Within the active annulus, the nematic director still has a chiral orientation and, as a result, the ring of active material rotates (Movie 3).
 \\

Fig.~S4 in the Supplemental Material \cite{smnejad} compares contributions to the droplet free energy as the activity is increased. The figure shows that droplet elongation and the separation of the two $+1/2$ defects as the activity increases leads to a decrease in the elastic energy. The surface tension energy remains almost unchanged, but shows a sudden jump when the hole forms.
The decrease in the elastic free energy density is not large enough to compete with the increase in the surface free energy density; therefore the hole formation can be identified as an active process.

\noindent
{\it Hysteresis:}

To quantify the change in the shape of the droplet, in Fig.~\ref{fig2} (b) we define $\ell_1$ and $\ell_2$ ($r_1$ and $r_2$) as the smaller and larger outer (inner) radius of the droplet.
In Fig.~\ref{fig2} (c) we start the simulations from a circular droplet for each value of the surface tension and measure $\ell_1$ and $\ell_2$. Decreasing the surface tension, the droplet first elongates and then forms a hole. 
By contrast, in Fig.~\ref{fig2} (d) we start the simulations from an annulus droplet and measure $r_1$ and $r_2$ as a function of surface tension. As expected, for large values of surface tension, the annulus collapses back to a configuration with no hole. Decreasing the surface tension, the radius of the active annulus increases. 
The same sequence of transitions, from circle to ellipse, to the annulus, is seen as the activity or the drop radius is increased (Figs. S4 and S5 in the Supplemental Material \cite{smnejad}).

Figs.~\ref{fig2} (c) and (d) show that the system has considerable hysteresis, with the steady state depending strongly on the initial conditions. In particular, the elliptical configuration can be reached from an initial circle, but not from an annulus as the surface tension is increased.

In Fig.~\ref{fig3} we show similar hysteresis as a function of activity. We 
measured   $\ell_1$ and $\ell_2$  as a function of activity, starting from two different initial conditions, a circular droplet, and an annulus. 
The graph shows that there is a range of activity, $0.008<\zeta<0.01$, for which the steady state in the simulations preserves the topology of the initial condition, forming a rotating elongating droplet without, or with, a hole. For very large activity $\zeta \ge 0.01$, the circular droplets also create a hole.

\noindent
{\it Growing droplets:} 

The theories of active nematohydrodynamics have proved successful in modelling the dynamics of bacteria and epithelial cell layers. These are systems where cell division is relevant and therefore it is interesting to study the configurations of growing active droplets. We consider parameters where the time scale for the growth is much larger than the active time scale $t_g = 1/k_d > t_a = \eta/\zeta$ ($2\times 10^6 > 10^2$).
Without any temporal perturbations, the drop grows symmetrically and the growth suppresses rotation and hole formation.

However, this behaviour changes when we add a uniform noise in the concentration field of magnitude $|\delta \phi| < 0.1$ every $\Delta T= 30000$ time-steps.
Fig.~\ref{growth2d} shows snapshots from simulations of a growing droplet (see also Movie 4).  The initial radius of the droplet is sufficiently small  that the droplet is isotropic and inert at the beginning of the simulations ($t_1$). As the droplet grows, it forms a vortex at the center and starts rotating ($t_2$). As it becomes larger, it accommodates more than one internal vortex and the interface of the droplet starts deviating from a circle ($t_4$). When the droplet becomes larger, still active flows become strong enough to initiate the formation of arms that fold and form holes ($t_5$-$t_{12}$).

 In the opposite regime where the growth time scale is smaller than the active time scale, we do not see rotation and hole formation as the growth suppresses the rotational flows created by the activity.\\

\section{THREE DIMENSIONS}

\begin{figure*}[t] 
    \centering
    \includegraphics[width=0.99\textwidth]{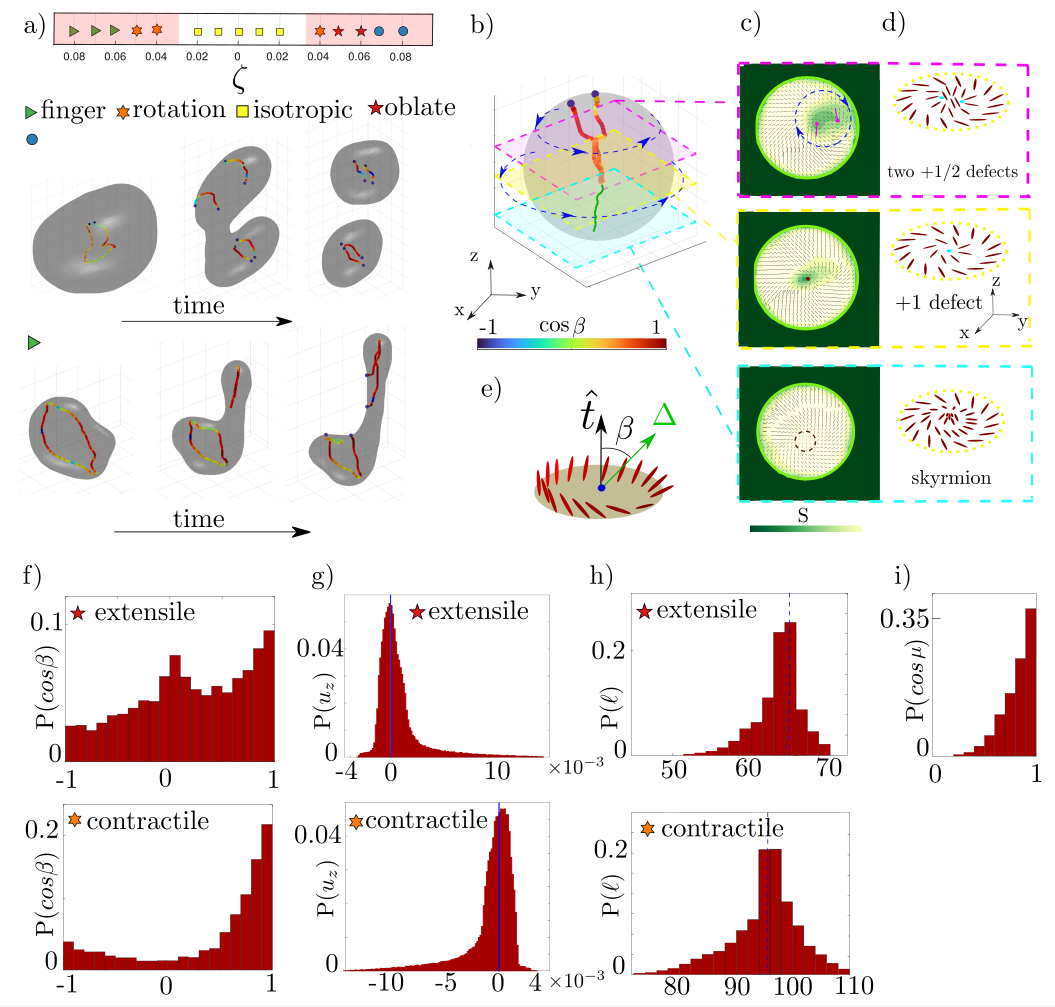}
    \caption{(a) Dynamical steady states of a 3D isotropic droplet. The pink background color indicates regions in which the droplet spontaneously rotates. Increasing the magnitude of the activity from zero leads to spontaneous rotation. A further increase in the magnitude of the activity in extensile droplets leads to the formation of an oblate ellipsoid. In contractile droplets, increasing activity leads to the formation of arms. Both extensile and contractile droplets break up at high activities (only shown for the extensile case here). A two-dimensional phase diagram as a function of activity $\zeta$ and surface tension coefficient $K_\phi$ is shown in Fig. S2(a) in the SI.
(b) Snapshot of a 3D rotating droplet with a stable disclination line. 
The blue dashed lines with arrows show the direction of rotation of the droplet. Disclination is colour coded by $\cos \beta$ where $\beta$ is the twist angle. The green line shows the skyrmion.
(c) Director field in different cross-sections along the disclination line. The color shows the magnitude of the nematic order.
(d) Schematic of the director field around the disclination line in the different cross-sections. The cyan points show the centre of the defects.
Due to the elastic interaction between different cross sections, the skyrmion has the same handedness as the disclination line.
(e) The vector around which the director rotates looping around a defect is shown by $\Delta$. The twist angle $\beta$ is defined as the angle between $\Delta$ and the vector $\hat{t}$ along the disclination line. 
 (f) Distribution of $\cos \beta$ in extensile (top) and contractile (bottom) droplets. Twist defects are more common in extensile systems.
(g) Breakdown of apolar symmetry due to the positioning of the disclination line on one side of the droplet leads to flows along the disclination line. The flows are along $+z$ in the extensile system and $-z$ in the contractile case.
(h) Distribution of the lengths of the disclination line in extensile and contractile systems. In extensile systems, the disclination lines are shorter, as the active flows are in the $+z$ direction. Here, we used $K_Q=0.05$, $K_{\phi}=0.02$, and $\ell_0=20$. (i) The distribution of $\cos \mu = \hat{t} \cdot \hat{\nu}$, where $\hat{\nu}$ shows the direction of the rotation axis. }
\label{fig:3da}
\end{figure*}

We next investigate whether spontaneous rotation can also be observed in 3D isotropic droplets. 
 Fig.~\ref{fig:3da}(a) shows how the steady states of the droplet vary for different values of activity.
Firstly we observe, as in 2D, that for very small values of activity, the isotropic phase is stable and the droplet does not show any collective behavior. 
Increasing activity, nematic order emerges and the droplet starts rotating (orange stars in Fig.~\ref{fig:3da}(a)) for both extensile and contractile driving. As it is clear from Eq. (\ref{dispersionb}), for the formation of the nematic order, we need $\zeta \lambda >0$, which means in extensile systems we have chosen $\zeta,\lambda>0$, and in contractile systems $\zeta, \lambda<0$. 

The rotation is caused by a disclination, which loops from the surface into the bulk and back to the surface, and which defines the axis of rotation (Fig.~\ref{fig:3da}(b)), which we take to be the $z$-axis, with the positive $z$ direction from the centre of the drop towards the ends of the defect line. Measuring the director field in cross sections perpendicular to the rotation axis shows two +1/2 defects close to the droplet surface which merge into a $+1$ defect further from the surface (Fig.~\ref{fig:3da}(c) and (d) top and middle row). Beyond the point where the loop terminates in the bulk, the  director configuration is a skyrmion (Fig.~\ref{fig:3da}(c) and (d) bottom row). The difference between a $+1$ defect and a skyrmion is that the magnitude of the order goes to zero at the center of the former, whereas it remains large and points out of the cross-sectional plane for the latter.

The details of the director field measured on planes perpendicular to dislocation lines have been shown to vary between extensile and contractile systems; twist-type configurations, where the director points out of the cross-sectional plane are more common in extensile systems \cite{PhysRevLett.128.048001,PhysRevResearch.5.L022061}. We quantify this in terms of the twist angle $\beta$, defined as the angle between the rotation vector $\Delta$, and the vector $\hat{t}$ tangent to the disclination line (Fig.~\ref{fig:3da}(e)). For +1/2 (-1/2) defects, $\Delta$ is parallel (antiparallel) to $\hat{t}$, and $\cos{\beta}=1$ ($\cos{\beta}=-1$). The out-of-plane twist component is maximal for $\cos{\beta}=0$. In Fig.~\ref{fig:3da}(f) we plot the distribution of  $\cos{\beta}$ in extensile and contractile droplets showing clearly that 
 twist sections of the disclination line are more common in the extensile case.

The disclination line is responsible for active forces and flows that differ in direction between extensile and contractile droplets. Recall, from Eq.~(\ref{eqn:active}), that the active force density is  $-\zeta \nabla \cdot \textbf{Q}$, where $\zeta>0$ ($\zeta<0$) in extensile (contractile) systems. Thus the force along the $z$-axis is $F_z=-\zeta \partial_z Q_{zz}$ (noting that the derivatives along perpendicular axes are zero by symmetry). 
As $Q_{zz}$ decreases along the $z$-direction, from a positive value at the center of the skyrmion to zero at the center of the defect, $F_z$ is positive (negative) in extensile (contractile) systems. 
Measuring the $z$-components of the flows (Fig.~\ref{fig:3da}(g)) indeed shows that large flows are along $+z$  in extensile droplets, and $-z$  in contractile droplets.
A consequence of this is that the disclination lines are shorter in extensile systems than in contractile systems (Fig.~\ref{fig:3da}(h)). 
In Fig.~S6 in the Supplemental Material \cite{smnejad} we display the average velocity field parallel and perpendicular to the disclination line inside the drop.

\begin{figure*}[t] 
    \centering
    \includegraphics[width=0.99\textwidth]{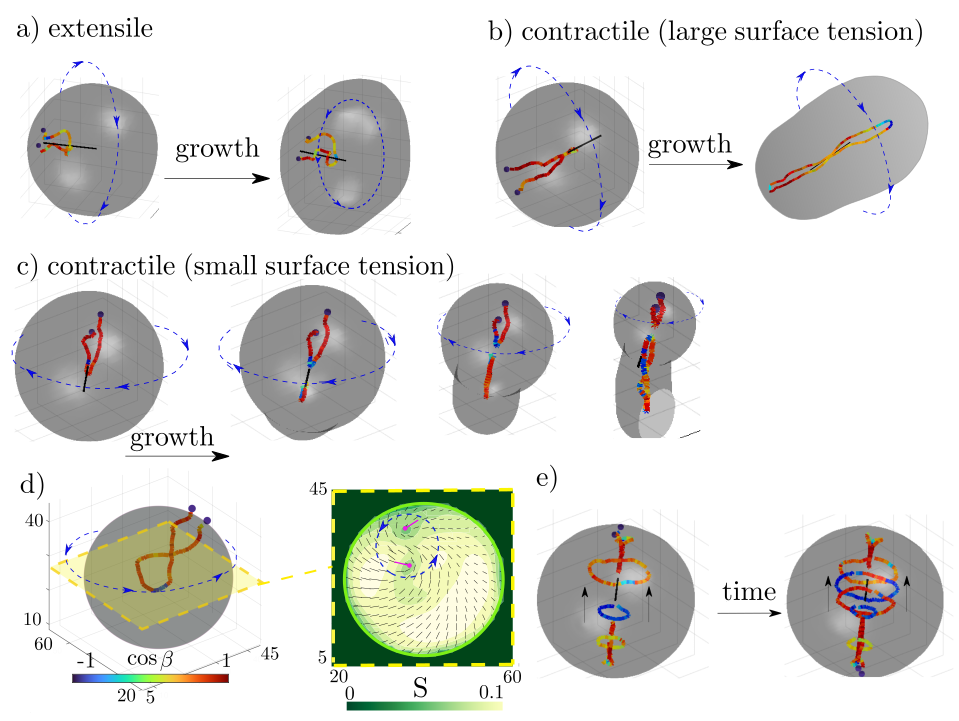}
    \caption{(a) A growing extensile droplet forms a rotating oblate ellipsoid. (b) A growing contractile droplet forms a rotating prolate ellipsoid. (c) For smaller values of surface tension ($K_\phi=0.01$), a contractile growing droplet forms an arm. (d) Snapshot of a  droplet where a wave is moving along the disclination line, and corresponding director conflguration in the yellow plane perpendicular to the disclination line showing two $+1/2$ defects which rotate around each other. The background color shows the magnitude of the nematic order.
 (e) Snapshot showing disclination loop creation. Disclination lines located at two sides of the droplet shoot disclination loops along its symmetry axis.
Disclinations are colour coded by $\cos \beta$ where $\beta$ is the twist angle. The blue dashed lines with arrows show the direction of rotation of the droplets and the black line shows the rotation axis. We used $K_{\phi}=0.01$ in panel (c). For all the other plots, we use a larger surface tension $K_{\phi}=0.02$. A phase diagram as a function of activity $\zeta$ and surface tension coefficient $K_\phi$ is shown in Fig. S2(b) in the SI.}
    \label{fig:3dd}
\end{figure*}

Stronger active flows alter the shape of the droplet. An initially spherical extensile droplet deforms to an oblate spheroid, contracting along the $z$-axis (red stars in  Fig.~\ref{fig:3da}(a)). This makes it easier for the two twist defects in the two arms of the dislocation line to separate. A further increase in extensile activity then leads to droplet break-up (blue circles in  Fig.~\ref{fig:3da}(a)).

By contrast, for contractile activity, the +1/2 defects are more stable \cite{PhysRevLett.128.048001} and the disclination line pushes the active fluid towards the skyrmion, and as a result, the droplet forms an arm that grows towards the skyrmion (green triangles in  Fig.~\ref{fig:3da}(a)). 

Similar shape changes are seen when we consider growing drops for which the growth time-scale $\tau_g=1/k_d=2\times 10^5$ is much larger than activity time-scale $\tau_a=\eta/\zeta\sim 10^2$. The droplet growth is illustrated in Fig.~\ref{fig:3dd}, and  Movies 5-7. As before, extensile droplets grow to oblates. Contractile droplets grow to prolates or, for lower surface tension, form an arm in the direction of the disclination line. The rotation and growth of the 3D contractile droplets is reminiscent of the unidirectional egg chamber elongation observed in \textit{Drosophila} \cite{cetera2014epithelial}. 
For growth which is very rapid compared to the active time scale no rotation or droplet shape changes are observed.

We now comment briefly on droplets that have a smaller value of the elastic constant ($K_Q = 0.02, K_{\phi}=0.02, 0.02<|\zeta| <0.04$).
 For these parameters, the disclination remains as a single line with cross sections that look like +1 defects in the plane perpendicular to the disclination line. The droplet shapes are more stable, and the rotation is accompanied by a self-propulsion due to the active flows along $z$. As expected extensile and contractile droplets move in opposite directions with respect to the orientation of the central disclination line. The more stable shape of the droplets is caused by the stability of the $+1$ disclination line which in a regime with larger elastic constant would form $+1/2$ and twist-type sections that lead to chaotic flows.

Moreover, in the contractile case for larger activities ($-0.05<\zeta \leq -0.04$) we observed a wave that propagates along the disclination line (Fig.~\ref{fig:3dd}(d) and Movie 8). This is caused by the rotation of the two $+1/2$ defects in the cross section around each other as also observed in 2D. In extensile systems, +1/2 defects are not stable, and twist regions in the defect loops suppress the wave. 

Finally, in this regime of small elasticity, for specific values of the parameters 
($K_{\phi}=0.06, -0.06<\zeta \leq -0.04$), we observed an unusual phase in which defect loops form and move toward the rotation axis. This is shown in Fig.~\ref{fig:3dd}(e).

\section{Lattice Boltzmann lengthscales and timescales}
In this section, we list lengthscales and the timescales that appear in the dynamics in Eqs. (\ref{eqn:lc}-\ref{eq10a}).
\begin{table}[ht]
\caption{Lengthscales and timescales of the equations (\ref{eqn:lc}-\ref{eqn:cont}).}\label{table1st}
\begin{tabular}{|c c|} 
 \hline
Length scales & Definition  \\ [0.5ex] 
 \hline\hline
 Defect length scale & $\ell_Q=\sqrt{K_Q/\mathcal{A}} \sim$  3 \\ 
 Nematic length scale& $\ell_c=\sqrt{K_Q/\mathcal{C}} \sim$  4 \\
 Active length scale  & $\ell_{\zeta}=\sqrt{K_Q \gamma \eta/(\lambda \zeta)} \sim$  3\\
 Interface length scale  & $\ell_{\phi}=\sqrt{K_{\phi}/A} \sim$ 1\\
 Advection length & $\ell_{\phi \zeta}=(\Gamma_{\phi} K_{\phi} \eta/(\zeta \lambda))^{\frac{1}{4}} \sim$ 2\\ [1ex] 
  \hline
Time scales & Definition \\
    \hline
 Active time scale  & $\tau_{a}=\eta/(\zeta \lambda) \sim$ $10^3$\\
 Growth time scale  & $\tau_{g}=1/k_d \sim$ $10^6$\\ [1ex] 
 \hline
\end{tabular} 
\end{table}

We assume that activity plays the dominant role in the stress, and that we can ignore the backflow terms that come from free energy. An approximate value for the velocity $u$ can then be found from the Navier-Stokes Eq. (\ref{eqn:cont}) in the low Reynolds number regime as $u \sim \zeta L/\eta$, where $L$ is a relevant length scale in the problem. Using this as an approximation for the velocity, we can find other relevant lenghscales and timescales as presented in Table \ref{table1st}.

It is apparent that there is no separation of length-scales, and as a result, the phase diagram cannot be explained in terms of one or two dimensionless numbers. There is however, a separation between the growth timescale $\tau_g$ and the active timescale $\tau_a$ as $\tau_g \gg \tau_a$. As a result, the role of the growth in the dynamics is to add extra material over long times thus making it possible for the active flows to form instabilities.

Although parameters that appear in Table \ref{table1st} are of the same order of magnitude, many of these parameters disappear in the linearised equations, and the formation of the nematic order due to the activity, can be predicted by a dimensionless number found in Eq.~(\ref{dispersionb}). In an infinitely large system ($q \sim 0$), the dimensionless number is equal to $\zeta \lambda/(\mathcal{A} \gamma \eta)$. 

\section{Comparing parameter values to experiments}

The nematohydrodynamic equations follow from symmetry arguments and obtaining quantitative values for the equation parameters for any given experimental system is challenging.
However, recent machine-learning approaches have started to investigate the dynamics of living systems using coupled partial differential equations \cite{PhysRevLett.129.258001,zhou2021machine,colen2021machine,PhysRevLett.129.258001}. In Ref. \cite{PhysRevLett.129.258001} a machine-learning approach was used to predict the equations of the nematic tensor and velocity field in an active microtubule motor protein suspension from experimental data. The parameters they found showed that the dynamics can indeed be described by the nematohydrodynamic equations at low Reynolds number with a positive activity $\zeta>0$, a non-zero flow-aligning parameter $\lambda>0$, and with an isotropic ground state, as in the model we use here. 
 They also found the ratio between activity and viscosity for different concentrations of ATP. A comparison between the parameters in our simulations and those extracted from the experiments is given in Table \ref{table2nd}.

The parameters used in the simulations are in lattice Boltzmann units. 
To map to physical units, we need a physical reference scale for three independent lattice Boltzmann parameters, such as
lattice spacing $\Delta x$, time step $\Delta t$, and viscosity $\eta$ \cite{kruger2017lattice}. Table \ref{table1st} suggests that if we choose $\Delta t$ to be in the interval $0.002 s \lessapprox \Delta t \lessapprox 0.01 s$, our system is in the same range of parameters as in the experiments. For this choice of the time step $\Delta t$, the total time of our simulations is between 2 $-$ 9 hours.

Another recent paper has used compliant elastic inclusions in an active tubulin-kinesin suspension to directly measure the magnitude of the activity, viscosity, and elastic constant to be $\zeta \sim (0.2 - 2) \times 10^{-7}$ Pa m,  $\eta\sim (4 - 10) \times 10^{-6}$ Pa s m, and $K_Q \sim (4 - 10) \times 10^{-16}$ N m, respectively \cite{Probingactive6}. Using a temporal grid $\Delta t \sim 0.01 s$, a force scale $\Delta F \sim 100 p N$, and a spatial grid scale $\Delta x \sim 10 \mu m$, the parameters used in our simulations are $\zeta \sim  10^{-7}$ Pa m, $\eta\sim  10^{-6}$ Pa s m, and $K_Q \sim 10^{-16}$ - $10^{-17}$ N m which matches well with the measurements in these experiments. 
\begin{table}[hb]
\caption{Comparing parameter values to experiments.}\label{table2nd}
\begin{tabular}{|c c c |} 
 \hline
Variable & Simulations & Microtubule suspension \\ [0.5ex] 
 \hline\hline
 $\rho u L /\eta$   & $10^{-3}$ $-$ $10^{-4}$ & 0  \\
 $\lambda$  & 0.7 & 0.37 $-$ 0.93  \\
  $\zeta/\eta$ & $10^{-3}$/$\Delta t$& (0.02 $-$ 0.5) $s^{-1}$\\ 
 $\gamma \mathcal{A}$ & $3\times 10^{-4}/\Delta t$ & (0.04 $-$ 0.16) $s^{-1}$\\[1ex] 
 \hline
\end{tabular}
\end{table}

Given this mapping to physical units our simulations show that rotation can appear in droplets with $\ell_0< 70 LB = 700 \mu m$. For larger droplets, when surface tension is sufficiently large to keep the droplet shape stable, active turbulence appears and suppresses the spontaneous rotation. Indeed, experiments on active microtubule kinesin motor suspensions confined to a circular boundary have reported a similar critical radius of $l_0 \sim 800 \mu m$ above which spontaneous rotation stops and active turbulence forms \cite{opathalage2019self}. 

Since spontaneous rotation has been observed in many experiments on epithelial colonies, we now compare our parameters with relevant measurements in epithelial systems. In Ref. \cite{blanch2017effective} a viscous active fluid model, similar to ours, along with simultaneous measurements of local stresses and velocities of an epithelial layer 
has been used to estimate the shear viscosity $\eta \sim (1-10)$ Pa s m. A typical order of magnitude of the contractile stresses for a 2D actomyosin system is of the order of $\zeta \sim 5\times 10^{-3}$ Pa m \cite{RevModPhys.85.1143}, for a thickness about the cell size (i.e. $h \sim 5 \mu m$). This value is comparable to the typical traction
stress of adhesive cells \cite{du2005force}. Assuming similar values for epithelial layers, the activity to shear viscosity ratio of epithelial layers can be estimated as $\zeta/\eta \sim 10^{-4}-10^{-3} s^{-1}$. The effect of the elasticity of the epithelial cells in the dynamics of their orientation has been estimated in Ref. \cite{lee2011crawling} to be of the order of $\gamma K_Q \sim 10^{-14} m^2/s$. A mapping between the lattice Boltzmann units and epithelial cell colonies ($20 s<\Delta t <200 s$, $10 \mu m<\Delta x <28 \mu m$, $5\times 10^{-6} N<\Delta F<1.4 \times 10^{-5}N$) leads to a critical radius $700 \mu m< \ell^c_0 < 2 \times 10^3 \mu m$ above which spontaneous rotation stops. These calculations show that epithelial cells can rotate in much larger circular colonies than microtubule kinesin motor droplets. In agreement with this result, spontaneous rotation has been observed
in gigantic millimeter-scale epithelial colonies in Ref. \cite{heinrich2020size}.

\section{Discussion}
We have solved the equations of active nematohydrodynamics to investigate the spontaneous rotation of unconfined droplets in both two and three dimensions. We work in a regime where the passive material has an isotropic director field, and nematic ordering is a consequence of active flows. To check this in experiments, one can measure the magnitude of the nematic order in the absence of activity. Indeed, it has been observed that activity can lead to the formation of nematic order in otherwise isotropic cell monolayers \cite{alltogether,doo2018}. Formation of the nematic order by active flows has also been observed in microtubule and kinesin suspensions \cite{hardouin2019reconfigurable}. 

The activity leads to spontaneous rotation in 2D droplets.  Using linear stability analysis, we show that the critical activity needed for the formation of order and collective motion scales linearly with $K_Q/\ell_0^2$, where $K_Q$ and $\ell_0$ are the elastic constant and the radius of the droplet, respectively. The rotation is caused by a chiral $+1$ defect, which can split into a chiral configuration of two $+1/2$ defects \cite{opathalage2019self,PhysRevLett.109.168105}. Decreasing surface tension or increasing activity leads to droplet elongation, and then hole formation and a rotating annulus. The parameters for which the hole forms depend strongly on the initial conditions. Our simulations predict that typically a droplet with surface tension $ k_\phi \sim 10^{-17}-10^{-16}$ N m, activity $\zeta \sim 10^{-7}$ Pa m, and flow-aligning parameter $\lambda \sim 0.7$,  becomes unstable and forms a hole when the radius of the droplet is $\ell_0 \sim 700 \mu$m.

Many experiments have observed the rotation of small cell colonies. At confluent but low densities, confined C2C12 myoblasts self-organized into spiral defect configurations with persistent rotation \cite{guillamat2022integer}. Heinrich {\it et al.\ }\cite{heinrich2020size} observed rotation in circular colonies, also of epithelial cells with a spiral +1 defect at the centre; in this paper, a polar model was used to explain the experimental data. Formation of spiral $+1$ defects and rotation has also been observed in sea urchin eggs where microtubules order to form spiral $+1$ defects which drive cytoplasmic flows \cite{schroeder1985spiral,suzuki2017spatial}. Rotation has also been observed in confined confluent MDCK epithelial cells \cite{doxzen2013guidance}, and in growing colonies of T84 cells, as long as the colonies were small and approximately circular \cite{ascione2022collective}. In the latter case, the results were interpreted in terms of an active nematic model. 

In addition to cell and bacterial colonies, rotation can also be observed in biofilaments and motor protein suspensions \cite{PhysRevLett.127.148001,berezney2022extensile,schroeder1985spiral,ndlec1997self}. In particular, in agreement with our results, in Ref. \cite{joshi2023disks} a microtubule and kinesin motor suspension inside an annulus was studied and it was shown that when the anchoring strength is low, microtubules form a spiral configuration with a non-zero but constant velocity. We predict that starting from a low motor concentration and increasing the concentration could lead to a transition between an isotropic phase where microtubules point in random direction, to a spiral configuration with a director that has an angle $\theta \sim 45^{\circ}$ from the radial direction, and finally to a configuration with $\theta > 45^{\circ}$ (see Fig. S3). 

Finally, the formation of a hole and rotating annulus has also been observed in bacterial suspensions in low concentration where the passive interaction (excluded volume) does not lead to nematic order \cite{nakane2021large}. Since bacteria are extensile ($\zeta>0$) and elongated ($\lambda>0$), it is very likely that the formation of rotation and hole is caused by the same mechanism we introduced here. This could be confirmed by measuring the activity, elastic constant, surface tension, and droplet size in these experiments, and comparing them with our values.

Our simulations predict that the activity also leads to the formation of nematic order and unidirectional rotation in three dimensions. In 3D, a disclination line is formed which
loops from the surface into the bulk and back to the surface, and defines the axis of rotation. Beyond the point where the disclination loops back, the rotation axis is marked by a skyrmion.
\\
In extensile systems, the disclination line has large twist sections, and cross sections with $+1$ defect configurations are less favored. As a result, the two arms of the disclination line increase their distance apart and the droplet elongates perpendicular to the disclination line. In contractile systems, on the other hand, $+1/2$ defects in disclination line cross-sections are more stable and the disclination line has stable sections with charge $+1$  in the planes perpendicular to the disclination line. 
We showed that activity leads to formation of flows along the disclination line. 
These tend to lengthen the disclination line in contractile systems and shrink it in extensile systems. As a result of this, and the different director configurations, extensile droplets tend to form oblate ellipsoids whereas contractile droplets tend to elongate along the rotation axis. 

In 3D, spontaneous persistent rotation has been observed in several biological contexts. Small spherical tissues rotate, an effect often termed coherent angular motion in the literature. Palamidessi {\it et al.} \cite{palamidessi2019unjamming} observed rotation in cancer spheroids of radius $\sim$ 100 $\mu$m and Brandstatter {\it et al.} \cite{brandstatter2021curvature} in spheroids of similar dimensions consisting of human mammary cells.

Persistent rotation has also been observed in several experiments on epithelial spheres.
Epithelial gland morphogenesis, where cylindrical branches transform into spherical alveoli during growth, is a morphogenetic transition that is accompanied by rotation of the emerging alveoli \cite{fernandez2021surface}.    
Epithelial spheres in pancreas-derived organoids also exhibit persistent rotation. Similarly to our model, the rotation is caused by spontaneous chiral symmetry breaking \cite{tan2022emergent} suggesting the absence of boundary-imposed topological charge.

When MDCK cells proliferate in a gel, they first form a spherical cluster, and then a cyst that consists of cells at the boundary of the sphere \cite{barlan2017fat2,hirata2018active}.
Active K-RAS induces rotation in the cell cluster, and $\beta$-catenin plays an important role in the rotation of the cyst. 
The epithelial cell layer that lines the Drosophila egg chamber also rotates during development \cite{cetera2014epithelial}, possibly to align fibril-like structures in the basement membrane surrounding the egg and to promote egg elongation. In Ref. \cite{barlan2017fat2} it is shown that Fat2 and Lar signaling promotes epithelial motility and leads to rotation in the Drosophila follicular epithelium. To compare the mechanism of the rotation with our simulations, it would be interesting to study cell elongation and nematic order in the presence and absence of these signals.

There are still many questions about both the physical description of the rotations, and about possible biological advantages they confer. It would be of interest to look experimentally for the changes in shape and hole formation in 2D colonies, predicted here, which are most likely to be observed in highly active cellular systems or active filaments and bacterial suspensions that form an isotropic phase in the absence of activity. Our numerical results suggest that it is easier to observe a stable hole if the experiments start with a circular colony (or suspension of active filaments/bacteria) that initially has a hole, as in this regime, the stable hole is accessible over a larger parameter span. In 3D experiments with rotating colonies, it would be interesting to measure and compare experimental values of activity, elasticity, and flow aligning parameter with the ones predicted by our model. 

\section*{Acknowledgements}
M.R.N. acknowledges the support of the Clarendon Fund Scholarships. 
\bibliographystyle{apsrev4-1}
\bibliography{references}
\end{document}